\documentclass[submission,copyright,creativecommons]{eptcs}
\providecommand{\event}{SYNT 2016} 
\usepackage{breakurl}             

\usepackage{amsfonts}
\usepackage{comment}
\usepackage{listings}

\usepackage{xcolor}
\usepackage[normalem]{ulem}
\usepackage{amsmath}
\usepackage{graphicx}
\usepackage{hyperref}
\usepackage{subcaption} 
\usepackage{caption}
\makeatletter
\ifcase \@ptsize \relax
  \newcommand{\miniscule}{\@setfontsize\miniscule{4}{5}}
\or
  \newcommand{\miniscule}{\@setfontsize\miniscule{5}{6}}
\or
  \newcommand{\miniscule}{\@setfontsize\miniscule{5}{6}}
\fi
\makeatother

\definecolor{lightgray}{rgb}{.9,.9,.9}
\definecolor{darkgray}{rgb}{.4,.4,.4}
\definecolor{purple}{rgb}{0.65, 0.12, 0.82}

\newcommand\Small{\fontsize{6.5}{9.5}\selectfont}

\lstdefinelanguage{scheme}{
  keywords={synth-fun,Int,Bool},
  keywordstyle=\color{blue}\bfseries,
  ndkeywords={synth-fun,Int,Bool},
  ndkeywordstyle=\color{purple}\bfseries,
  identifierstyle=\color{black},
  escapeinside={@}{@},
  numbers=left,
  numbersep=5pt, 
  numberstyle=\tiny,
  sensitive=false,
  comment=[l]{;},
  commentstyle=\color{purple}\ttfamily,
  stringstyle=\color{red}\ttfamily,
  morestring=[b]',
  morestring=[b]"
}

\makeatletter
\def\url@smallstyle{%
  \@ifundefined{selectfont}{\def\UrlFont{\sf}}{\def\UrlFont{\small\ttfamily}}}
\makeatother
\def\authorrunning{S. Chasins \& J. Newcomb}
\def\titlerunning{Using SyGuS to Synthesize Reactive Motion Plans}

\title{Using SyGuS to Synthesize Reactive Motion Plans}

\author{Sarah Chasins
\institute{Department of Computer Science \\ UC Berkeley \\ California, USA}
\email{schasins@cs.berkeley.edu}
\and
Julie L. Newcomb
\institute{Department of Computer Science \\ University of Washington \\ Washington, USA}
\email{newcombj@cs.washington.edu}
}

\begin{document}

\date{}

\maketitle

\begin{abstract}

We present an approach for synthesizing reactive robot motion plans, based on compilation to Syntax-Guided Synthesis (SyGuS) specifications.  Our method reduces the motion planning problem to the problem of synthesizing a function that can choose the next robot action in response to the current state of the system.  This technique offers reactivity not by generating new motion plans throughout deployment, but by synthesizing a single program that causes the robot to reach its target from any system state that is consistent with the system model.  This approach allows our tool to handle environments with adversarial obstacles.   This work represents the first use of the SyGuS formalism to solve robot motion planning problems.  We investigate whether using SyGuS for a bounded two-player reachability game is practical at this point in time.  

\end{abstract}

\section{Introduction}

Many robots operate in dynamic, nondeterministic environments.  Although a variety of techniques have been developed for synthesizing straight-line motion plans for robots in highly predictable environments, it is often difficult to extend these approaches to robots that must react to sensor values.  In particular, exciting new work on using off-the-shelf SMT solvers to produce robot motion plans \cite{saha2014,saha2016,nedunuri2014smt} has not thus far been extended to the reactive motion planning problem.

In contrast to a standard motion plan, a \emph{reactive} motion plan responds to changes in the environment.  Typically this means using sensors to observe the environment, and conditioning actions on the observed sensor values.  The ability to respond to a robot's surroundings makes reactive motion plans well-suited to changing environments, and especially to environments with adversarial agents.  

In this paper, we present a compiler from high-level descriptions of reactive motion planning problems to Syntax-Guided Synthesis (SyGuS) specifications, allowing us to generate reactive motion plans using off-the-shelf SyGuS solvers.  We focus specifically on discretized environments with adversarial moving obstacles.  The goal is to generate a correct-by-construction motion plan for a given workspace and a given set of moving obstacles.

\begin{figure}[t]
    \centering
    \begin{subfigure}[t]{0.46\textwidth}
        \centering
          \includegraphics[page=1, viewport=0 0 270 370, clip=true, width=.6\columnwidth]{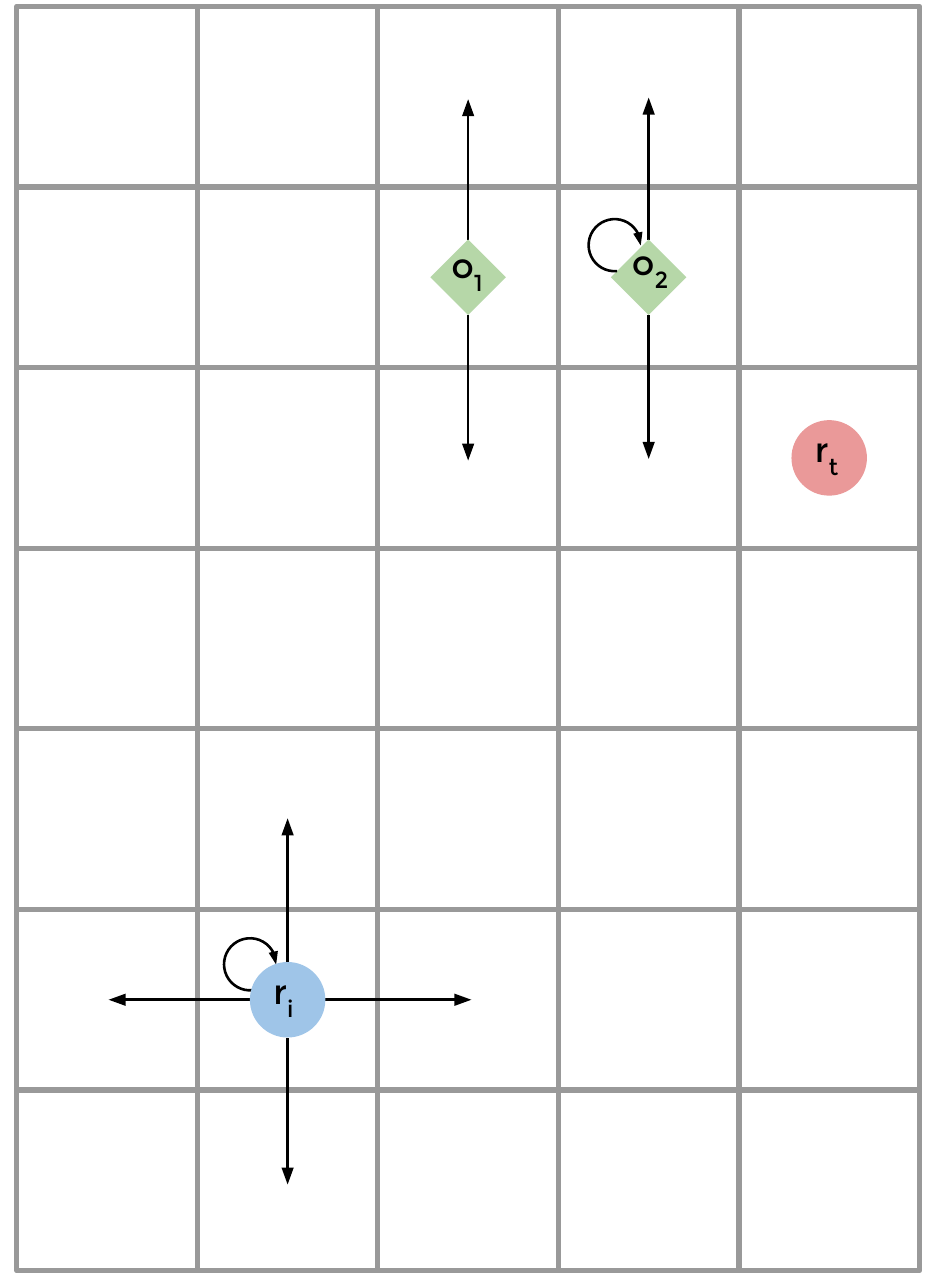}
            \caption{A pictorial representation of an input problem for our tool.  The blue $r_i$ circle represents the initial position of the robot.  The red $r_t$ circle represents the target position for the robot.  The green $o_1$ and $o_2$ diamonds represent the initial positions of the two obstacles.  The arrows extending from the initial positions of each agent represent the motion primitives available to the given agent.}\label{fig:sampleProblem}
    \end{subfigure}%
    \hspace{15pt}
    \begin{subfigure}[t]{0.46\textwidth}
        \centering
          \includegraphics[page=2, viewport=0 0 270 370, clip=true, width=.6\columnwidth]{sampleProblem.pdf}
        \caption{A pictorial representation of one solution to the problem posed in Figure \ref{fig:sampleProblem}.  Each arrow represents the application of a motion primitive.}\label{fig:sampleProblemSolution}
    \end{subfigure}
    \caption{Pictorial representations of an input to our tool, and a candidate solution.}
\end{figure}

More concretely, we are interested in generating motion plans for problems like the one portrayed in Figure \ref{fig:sampleProblem}.  Figure \ref{fig:sampleProblem} depicts a robot's initial starting position, target final position, and its available motion primitives (the set of moves it may make).  It also depicts two obstacles, each with its own set of available motion primitives.  Figure \ref{fig:sampleProblemSolution} shows a motion plan that meets the problem specification.  The goal of our work is to generate such motion plans automatically.

\section{Overview}

\label{sec:approach}

Our approach to the reactive obstacle avoidance problem is to accept as input exactly the type of information displayed in Figure \ref{fig:sampleProblem} and compile the motion planning problem to a SyGuS specification.  SyGuS offers a clean language for program synthesis \cite{sygusSpec,sygus,sygusExtended}.  It allows users to constrain synthesized programs both via grammars, and via constraints on the outputs.  The ease with which SyGuS allows users to explore programs with differing abstract syntax trees makes it a natural fit for the reactive motion planning problem, which may demand different control flow for different input problems.  While a space of straight-line programs of a given length can be naturally expressed with SMT, the standard SMT encoding essentially demands that all candidate programs share an AST.  In contrast, SyGuS is explicitly designed to facilitate exploring and comparing many program structures, as defined by a grammar.  Since a reactive motion plan may need at any point to react to changing environment conditions, custom control flow is central to the problem; thus, straight-line SMT encodings cannot be adapted to the problem without building a SyGuS-like layer on top of them.

We use SyGuS to generate a function that selects a motion primitive from among the available set, based on the current state of the system.  The function accepts the current position of the robot and the current position of all obstacles as inputs.  The final motion plan applies the function to the current state to select a motion primitive, then applies the returned motion primitive, then repeats these steps until the robot reaches the target position.
\\

\noindent We briefly emphasize a few important characteristics of our tool.

\paragraph{Synthesize-once.}  Our tool completes all synthesis before deployment.  In contrast to receding horizon techniques, our approach does not deploy the synthesizer with the robot or resynthesize the controller throughout execution.  Rather our tool generates, at design time, a single controller that can respond to all allowable environment behaviors.

\paragraph{Motion primitive agnostic.} Our synthesis approach, like many previous approaches \cite{saha2014,temporalLogic,motionPrimitives} decouples the generation of motion primitives from the selection of which motion primitive to use at any given step.  Our approach takes a set of pre-generated motion primitives as input, so it can be fruitfully combined with any motion primitive generation strategy whose output primitives can be expressed in our target format.  We must be able to extract the set of positions relative to the starting point through which a robot executing the primitive may pass, in the discretized space.

\paragraph{Relies on correct system model.}  Our work focuses on generating a controller that satisfies the specification for all executions described by the system model, rather than by the system itself.  Like past work on reactive motion plan synthesis \cite{recedingHorizon,temporalLogic}, our tool does not offer any guarantees of success if the environment breaks the assumptions imposed by the model.  Further, our work is not concerned with verifying that the model accurately describes the actual system.  Like past reactive synthesis tools, we consider that validation problem to be outside the scope of our work.

\paragraph{Does not require a custom solver.}  Although substantial work has gone into synthesizing non-reactive motion planners with off-the-shelf solvers, especially SMT solvers \cite{saha2014,nedunuri2014smt,saha2016}, work on synthesizing \emph{reactive} motion planners with off-the-shelf solvers is still in the early stages.  To our knowledge, this work is the first attempt to solve reactive motion planning problems by direct compilation to a \emph{functional} solver \footnote{We draw the same distinction between functional and reactive synthesis described by \cite{functional} and use the term \emph{functional solver} to refer to solvers that perform functional synthesis.}.  Because we use a functional solver, we achieve more control over the shape of output motion plans than we could exert using existing reactive solvers.

\section{SyGuS for Reactive Motion Planning}

In this section, we offer a formal model of our reactive motion planning problem.  We describe our approach for compiling an instance of the motion planning problem to a SyGuS specification.  Finally, we give an illustrative example.

\subsection{Formal Model of Our Reactive Motion Planning Problem}

\label{sec:formalModel}

Each instance of our motion planning problem is defined by information about the robot and its environment.

A motion primitive $M$ is a 2-tuple: $\langle p_{rf}, P_{r} \rangle$.  The relative final position after applying the motion primitive is $p_{rf}$.  That is, given a current position $\langle x,y \rangle$, the final position after applying $M$ is $\langle x+p_{rf}.x, y+p_{rf}.y \rangle$.  Every $p_r \in P_r$ is an intermediate relative position through which the robot may pass during execution of $M$.  \footnote{As a concrete example to further clarify the role of $P_r$, consider that we may have multiple motion primitives that result in a position one unit up and one unit right.  Perhaps one primitive causes the actor to travel up then right and another causes it to travel right then up.  Since the intermediate positions associated with these actions would differ, the motion primitives would differ.  These primitives would be $\langle \langle 1,1 \rangle, \{ \langle 0,1 \rangle \} \rangle$ and $\langle \langle 1,1 \rangle, \{ \langle 1,0 \rangle \} \rangle$, respectively.}

An input problem instance $\mathcal{P} = \langle D, R, \mathcal{O}, l \rangle$.  $D$ represents the dimensions of the workspace, $R$ is the robot, $\mathcal{O}$ is the set of obstacles, and $l$ is a limit on the length of the path.

$D = \langle width, length \rangle$ represents the dimensions of the workspace, $width$ in the x direction, and $length$ in the y direction.

$R = \langle p_i, p_t, \mathfrak{M} \rangle$.  The robot's initial position is $p_i$, the robot's target final position is $p_t$, and $\mathfrak{M}$ is the set of robot motion primitives.

Each $O \in \mathcal{O}$ is a tuple: $\langle p_i, \mathfrak{M} \rangle$.  The obstacle's initial position is $p_i$, and $\mathfrak{M}$ is the set of the obstacle's motion primitives.

\subsection{Compilation to SyGuS Specifications}

As input, our compiler takes an input problem $\mathcal{P} = \langle D, R, \mathcal{O}, l \rangle$, as described above.  The limit $l$ on the length of the path bounds the number of motion primitive applications in the motion plan.  Although our compiler to SyGuS treats $l$ as input, in practice the user does not provide $l$.  Rather, the tool generates the appropriate $l$ by increasing $l$ until the problem becomes satisfiable.

The output of our tool is a function for choosing the next motion primitive to apply.  The function takes the current position of the robot and the current positions of all obstacles as input.  Applying the function to the appropriate arguments for the target number of steps ($l$) produces the motion plan.

\begin{figure}
\begin{lstlisting}[caption={The form of the grammar, given that there are $n$ obstacles and $m$ robot motion primitives, and $d$ is the length of the longest dimension of the workspace.},label={lst:grammar}]
(synth-fun move ((p-r Int) (p-o1 Int) ... (p-on Int)) Int
	((Start Int (
		MoveId
		(ite StartBool Start Start)
	))
	(MoveId Int (
		1
		...
		m
  ))
	(CondInt Int (
		(get-x p-r)
		(get-y p-r)
		(get-x p-o1)
		(get-y p-o1)
		...
		(get-x p-on)
		(get-y p-on)
		(+ CondInt CondInt)
		(- CondInt CondInt)
		-1
		0
		...
		d-1
	))
	(StartBool Bool ((and StartBool StartBool)
		(or  StartBool StartBool)
		(not StartBool)
		(<=  CondInt CondInt)
		(=   CondInt CondInt)
))))
\end{lstlisting}
\end{figure}

The grammar for the synthesized function is given in Listing \ref{lst:grammar}.  The synthesized function takes the current state of the system as input and as output produces the next motion primitive that the robot should apply.  The grammar depends on the input problem in several ways.  First, the number of obstacle position parameters depends on the number of obstacles.  Second, the candidate return values correspond to the motion primitives available to the robot.  Third, the return value may be conditioned on the position of the robot, and also on the position of any of the obstacles.  Finally, we reduce the depth of the programs that the SyGuS solver must consider by including constants up to the maximum coordinate in the workspace.

The constraint on the synthesized function takes the form $(or \ (not \ obstaclePositionsAllowable) \\ (and \ correctFinalPosition \ noIntersections)$.  Intuitively, this constraint requires that either an obstacle made an illegal move --- in which case our tool provides no guarantees --- or the robot arrives in the correct final position, without having collided with any obstacles.

We model the obstacle positions as universally quantified variables.  Thus, some assignments to the obstacle positions are not consistent with obstacles' motion primitives.  The $obstaclePositionsAllowable$ constraint requires that for each obstacle, for each step, there is a motion primitive associated with the obstacle that could produce the next position.  Let $n$ be the number of obstacles, let $l$ be the number of steps the robot takes, let $o_{i,j}$ be the position of the $i^{th}$ obstacle ($O_i$) at time $j$, and let $\mathfrak{M}_i$ be the motion primitives of $O_i$.  Then the  $obstaclePositionsAllowable$ constraint is: \\ $\bigwedge\limits_{i=1}^{n}[ \bigwedge\limits_{j=1}^{l} [\exists M \in \mathfrak{M}_i . (o_{i,j} - o_{i,j-1} = M.p_{rf})]]$  

The $correctFinalPosition$ constraint requires only that the robot's final position after $l$ steps be the target position $p_t$ specified in $R$.  Because the synthesized function takes the current robot position as an input, we find the final position by applying the synthesized function to its own output $l$ times.  Let $move$ be the synthesized function for selecting motion primitives.  Then $r_j$, the position of the robot at time $j$, is $move(r_{j-1},o_{1,j-1},o_{2,j-1},...,o_{n,j-1})+r_{j-1}$.  Then the $correctFinalPosition$ constraint is simply: $r_l = R.p_t$.

The $noIntersection$ constraint requires that, for each step, there is no intersection between the positions visited during the robot's motion primitive and the motion primitives of the obstacles.  Because the step is the finest-grain measure of time in our system, both the robot and the obstacles may be in any location along their respective paths at any point during the step, including both the initial and the final location.  Thus, our $noIntersection$ constraint must require that the two paths are nonoverlapping.  For convenience, we now refer to the robot's motion primitive at step $j$ as $move_j$ and the $i^{th}$ obstacle's motion primitive at step $j$ as $omove_{i,j}$.  The $noIntersection$ constraint is: \\ $\bigwedge\limits_{i=1}^{n}[ \bigwedge\limits_{j=1}^{l} [\forall m \in (\{\langle 0,0 \rangle, move_j.p_{rf}\} \cup move_j.P_r) [ \forall m_o \in (\{\langle 0,0 \rangle, omove_{i,j}.p_{rf}\} \cup omove_{i,j}.P_r)$ \\ $[r_{j-1}+m \neq  o_{i,j-1}+m_o]]]]$

\subsection{Concrete Example}

\label{sec:concreteExample}

For a more concrete example, we turn to the sample problem discussed in Section \ref{sec:approach} and pictured in Figure \ref{fig:sampleProblem}.  To use our tool to solve that problem, the user would provide the following inputs:

\begin{itemize}

\item The dimensions of the workspace: $D = \langle 5, 7 \rangle$.

\item The robot's initial position, target position, and motion primitives: $R = \langle \langle1,1\rangle,\langle4,4\rangle, \\ \{\langle \langle0,0\rangle, \{\} \rangle, \langle \langle0,1\rangle, \{\} \rangle , \langle \langle1,0\rangle, \{\} \rangle, \langle \langle0,-1\rangle, \{\} \rangle, \\ \langle \langle-1,0\rangle, \{\} \rangle \} \rangle$.

\item The obstacles' initial positions and motion primitives: $\mathcal{O} = \{ \langle \langle2,5\rangle, \{\langle \langle0,1\rangle, \{\} \rangle, \langle \langle0,-1\rangle, \{\} \rangle \} \rangle, \\ \langle \langle3,5\rangle, \{ \langle \langle0,0\rangle, \{\} \rangle, \langle \langle0,1\rangle, \{\} \rangle, \langle \langle0,-1\rangle, \{\} \rangle \} \rangle \}$

\end{itemize}

Given this input, our tool would produce the grammar in Listing \ref{lst:grammarConcrete} as part of the SyGuS specification.  We also include the full SyGuS specification generated for this problem in Appendix \ref{appendix:fullSpec}.  Running our tool on this problem produces the output in Listing \ref{lst:solutionConcrete}.

\begin{figure}
\begin{lstlisting}[caption={The grammar for a concrete example.},label={lst:grammarConcrete}]
(synth-fun move ((p-r Int) (p-o0 Int) (p-o1 Int)) Int
	((Start Int (
		MoveId
		(ite StartBool Start Start)
	))
  (MoveId Int (
    0 ;corresponds to no move
		1 ;corresponds to up
		2 ;corresponds to right
		3 ;corresponds to down
		4 ;corresponds to left
	))
	(CondInt Int (
		(get-x p-r) ;x coord of robot
		(get-y p-r) ;y coord of robot
		(get-x p-o0) ;x coord of first obstacle
		(get-y p-o0) ;y coord of first obstacle
		(get-x p-o1)
		(get-y p-o1)
		(+ CondInt CondInt)
		(- CondInt CondInt)
		-1
		0 ;0-6 are possible coordinates in space
		1
		2
		3
		4
		5
		6
	))
	(StartBool Bool ((and StartBool StartBool)
		(or  StartBool StartBool)
		(not StartBool)
		(<=  CondInt CondInt)
		(=   CondInt CondInt)
))))
\end{lstlisting}
\end{figure}

\begin{figure}
\begin{lstlisting}[caption={The solution our tool synthesizes for the problem from Section \ref{sec:concreteExample}.}, label={lst:solutionConcrete}]
(define-fun move ((p-r Int) (p-o0 Int) (p-o1 Int)) Int
    (ite (<= (get-x p-r) 3) 2 1))
\end{lstlisting}
\end{figure}

\subsection{Implementation}

We have implemented the compiler described here for transforming environment descriptions and goals to the corresponding SyGuS specifications.  The source code of our tool is publicly available at: \\ \url{https://github.com/schasins/reactive-motion-planning-synthesis}.

\section{Evaluation}

In this section, we evaluate the scalability of our SyGuS-based approach to reactive motion planning.  We report how synthesis time scales with problem size, using several different measures of problem size.  We also discuss the choice of which SyGuS solver to use for these experiments. 

\subsection{Scalability}

To evaluate the scalability of our SyGuS-based approach, we consider how it scales over four axes of problem size: the depth of the AST to synthesize, the length of the path the robot must travel, the number of obstacles in the environment, and the dimensions of the environment.  We report performance results obtained using the {\sc Enumerative} SyGuS solver \cite{enumerative}, which won the SyGuS competition in 2014 \cite{sygusExtended}.  See Section \ref{sec:solver} for a discussion of the solver choice.

\subsubsection{AST Depth}

\begin{figure}
\centering
\includegraphics[page=1, viewport=40 470 490 725, clip=true, width=.6\columnwidth]{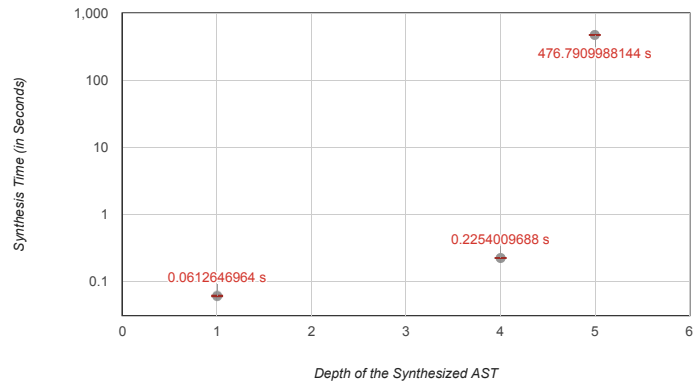}
\caption{Execution time of the SyGuS solver as a function of the depth of the AST it must synthesize.  For each benchmark measured here, the path length is 5, the number of obstacles is 4, and the environment size is 6 by 6.  Only the depth of the output AST varies.  Note that the synthesis time is shown on a logarithmic scale.}\label{fig:depth}
\end{figure}

We start by considering how our approach scales with the size of the synthesized programs, always a primary concern for synthesis techniques.  Specifically, we consider the relationship between solver execution time and the depth of the abstract syntax tree (AST) of the shallowest program that can satisfy the specification.  Figure \ref{fig:depth} shows the time to synthesize programs with depth 1, depth 4, and depth 5 ASTs.  Note that the structure of our generated grammars means that no reasonable program will have a depth of 2 or 3.  As we expect, our results reflect state space explosion.  The increase in synthesis time as AST depth rises is steep.

Although it is customary to compare a new synthesis tool's maximum program size with the maximum program size of previous synthesis tools, we have found no relevant synthesis tools for which we can compare AST depth.  It is somewhat difficult to compare a SyGuS-based synthesis tool with an SMT-based synthesis tool, since SMT-based tools typically measure the sizes of output programs in terms of number of instructions.  In our case, using SyGuS, our tool is synthesizing not only the `instructions' at each point in the program, but also the structure of the program itself.  Without a clean way of comparing measures of AST size to instruction counts, we defer the comparison with other motion planning synthesis tools to the section on path length, Section \ref{sec:steps}, where the number of steps in the robot motion path gives us a natural way of comparing program size.

The results of this experiment persuade us that as the size of the solution AST grows, the synthesis time grows very quickly.  However, we are encouraged that we can synthesize depth 5 ASTs -- already capable of expressing quite complex and sophisticated programs -- in less than 8 minutes.

\subsubsection{Path Length}

\label{sec:steps}

\begin{figure}
\centering
\includegraphics[page=1, viewport=40 470 490 725, clip=true, width=.6\columnwidth]{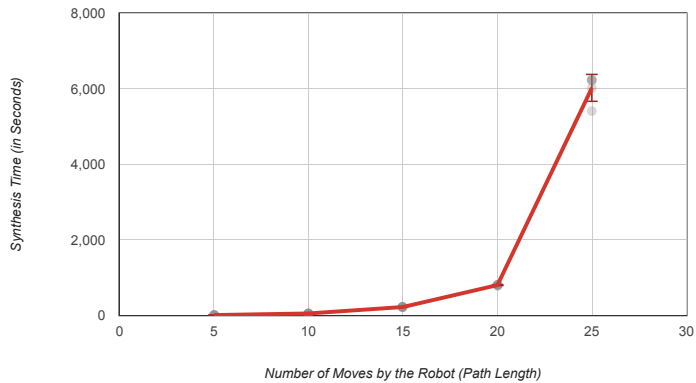}
\caption{Execution time of the SyGuS solver as a function of the number of steps (motion primitive applications) in the robot's path.  For each benchmark measured here, the AST depth is 4, the number of obstacles is 5, and the environment size is 20 by 20.  Only the length of the robot's path varies.}\label{fig:steps}
\end{figure}

Next we consider how the length of the robot's motion path affects synthesis time.  We measure the motion path in steps, where each step corresponds to one application of one of the robot's motion primitives.  Figure \ref{fig:steps} depicts the relationship between path length and synthesis time.  As we expect, longer paths require longer synthesis times.

While it is clear that a higher AST depth is associated with a larger program search space, it may be less immediately obvious how path length affects synthesis times.  We detail the primary mechanisms here.
\begin{itemize}
\item From the perspective of the SyGuS solver, a candidate program is a solution if the correctness condition holds for every possible combination of obstacle motion plans.  At each step, each obstacle chooses amongst its available motion primitives.  This means that at each step, if there are $n$ obstacles and obstacle $o_i$ has $p_i$ motion primitives, the number of possible obstacle outcomes at any given step is $\prod_{i=1}^{n} p_{i}$.  Thus, the space of obstacle moves across a whole path of length $l$ is $({\prod_{i=1}^{n} p_{i}})^l$.
\item For a fixed assignment of motion primitives to obstacle-step pairs, the solver must calculate the positions of the robot and the obstacles after each step.
\item For a fixed assignment of motion primitives to obstacle-step pairs, the solver must identify whether the robot collides with each obstacle at each step.
\end{itemize}

Although the curve is very steep, we are pleased that we are able to synthesize such long paths, given the usual limits on synthesized program sizes.

For comparison, the SMT-based motion plan synthesis tool Complan \cite{saha2014} was assessed on two case studies.  The authors report that they ceased searching for a solution to one case study after no path of length 20 or shorter satisfied the specification.  With this in mind, and given that our tool must synthesize control flow and handle obstacles moving at each step, we are pleased that our approach can reach that same range, and even beyond it.

Given the very great differences between the Complan motion planning tasks and our tasks, a comparison between the synthesis times of these tools is not very meaningful.  However, for the sake of arguing that our synthesis times are within an acceptable range, we note that Complan synthesizes a length 13 plan for one specification in 15 minutes and 36 seconds and a length 12 plan for a different specification in 25 minutes and 48 seconds.  For comparison, for the length 15 version of the Figure \ref{fig:steps} benchmark, our average synthesis time was 3 minutes and 37 seconds.  Again, these times cannot be meaningfully compared as measures of tool performance, and we do not intend to imply any serious comparison with Complan.  We include these numbers only as an indication that our synthesis times are within an acceptable range, relative to previous tools.  We believe this offers a strong indication that correct-by-construction synthesis of reactive motion plans using off-the-shelf solvers is feasible.

\subsubsection{Number of Obstacles}

\begin{figure}
\centering
\includegraphics[page=1, viewport=40 470 490 725, clip=true, width=.6\columnwidth]{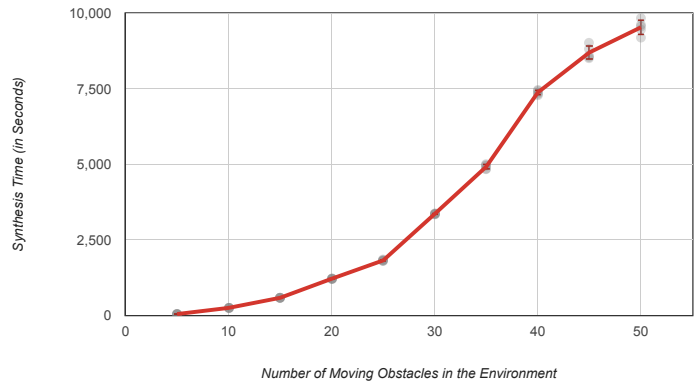}
\caption{Execution time of the SyGuS solver as a function of the number of moving obstacles in the environment.  For each benchmark measured here, the AST depth is 4, the path length is 5, and the environment size is 20 by 20.  Only the number of obstacles varies.}\label{fig:obstacles}
\end{figure}

Next we investigate the relationship between the number of moving obstacles in an environment and the synthesis time.  We show the synthesis times for a variety of different obstacle counts in Figure \ref{fig:obstacles}.  Naturally, increasing the number of obstacles increases the synthesis time.

The number of obstacles in the environment affects the synthesis time through the following mechanisms:
\begin{itemize}
\item The robot may react to the locations of obstacles, which are arguments to the synthesized function.  Thus, increasing the number of obstacles also increases the size of the grammar.
\item Recall that if there are $n$ obstacles, the space of obstacle moves across a whole path of length $l$ is $({\prod_{i=1}^{n} p_{i}})^l$.  Increasing the number of obstacles increases the number of obstacle-step variables, increasing the space the solver must explore.
\item For a fixed assignment of motion primitives to obstacle-step pairs, the solver must calculate the positions of each obstacle after each step.
\item For a fixed assignment of motion primitives to obstacle-step pairs, the solver must identify whether the robot collides with each obstacle at each step.
\end{itemize}

\subsubsection{Environment Dimensions}

\begin{figure}
\centering
\includegraphics[page=1, viewport=40 470 495 725, clip=true, width=.6\columnwidth]{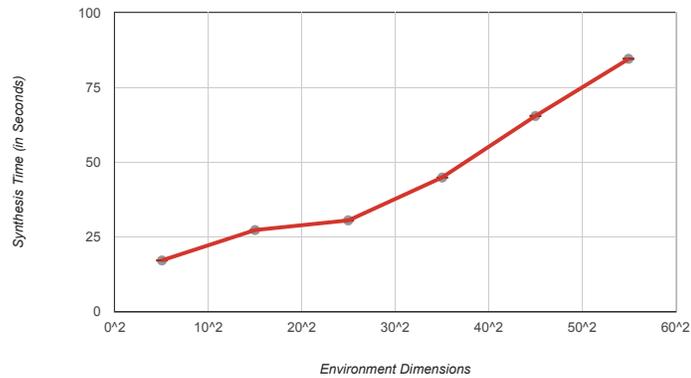}
\caption{Execution time of the SyGuS solver as a function of the dimensions (length and width) of the environment.  For each benchmark measured here, the AST depth is 4, the path length is 5, and the number of obstacles is 5.  Only the size of the environment varies.  The label ``50\^{}2'' indicates that both the length of the environment and the width of the environment were 50, for a total of 250 discrete positions in the environment.}\label{fig:dimensions}
\end{figure}

Finally, we consider the effect of environment dimensions on synthesis time.  Figure \ref{fig:dimensions} displays the relationship between the length and width of the discretized environment and performance.  Recall that we do include the range from zero to $max(length, width)$ in the output grammars, so increasing the dimensions does increase the size of the grammar.  Although increases in the environment dimensions do raise the synthesis time, the curve is not steep, so we consider our approach to scale well to large environments.

\subsection{Solver Selection}

\label{sec:solver}

Readers may be interested to note that although the {\sc Cvc4-1.5-sygus} placed first in the General Track of the 2015 SyGuS competition, we conducted our experiments using the winner of the 2014 SyGuS competition, the {\sc Enumerative} solver.   We chose the {\sc Enumerative} solver based on the results of the SyGuS 2015 competition \cite{sygus2015}, which included motion planning benchmarks generated by our tool.  The competition compared the performance on these benchmarks by all five of the 2015 General Track solvers: {\sc Cvc4-1.5-sygus}, {\sc Enumerative}, {\sc Stochastic}, {\sc Sketch-ac}, and {\sc SosyToast}.  Although {\sc Cvc4-1.5-sygus} placed first in the general track and {\sc Enumerative} placed only second, {\sc Enumerative} tended to outperform {\sc Cvc4-1.5-sygus} on benchmarks with multiple invocations.  Since repeated invocation of the synthesized function is a central feature of our approach, this alone would suggest the {\sc Enumerative} solver as a better fit.  However, we also have the benefit of observing how the solvers perform on our motion planning tasks specifically.  Of the solvers, only {\sc Cvc4-1.5-sygus}, {\sc Enumerative}, and {\sc Stochastic} solved any motion planning benchmarks.  {\sc Stochastic} only solved benchmarks produced by an earlier version of our tool; the earlier version produced SyGuS specifications that accommodated missing solver features but required longer synthesis times.  Thus, only {\sc Cvc4-1.5-sygus} and {\sc Enumerative} solved specifications produced by the current tool.  Of these, the {\sc Enumerative} solver completed 5 out of 6 of the tasks, while {\sc Cvc4-1.5-sygus} solved only 2.  In particular, {\sc Cvc4-1.5-sygus} solved only the benchmarks that combined short path lengths with small AST depths.  Even for the two benchmarks completed by both solvers, the {\sc Enumerative} solver exhibited faster synthesis times than {\sc Cvc4-1.5-sygus}.  With these results in hand, we are confident that the {\sc Enumerative} solver currently offers the best performance for our motion planning tasks.

\subsection{Discussion}

\begin{figure}
\centering
\includegraphics[page=1, viewport=40 470 490 725, clip=true, width=.5\columnwidth]{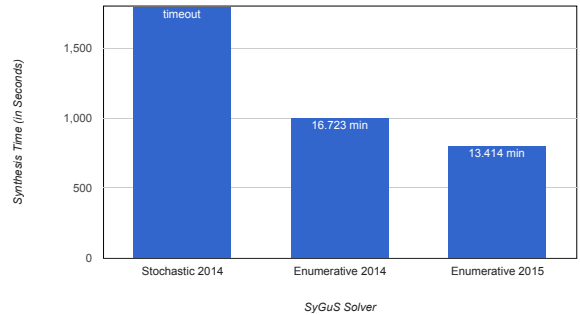}
\caption{Execution time of various SyGuS solvers on the same reactive motion planning problem, synthesizing a length 20 motion plan.  For this benchmark, the AST depth is 4, the path length is 20, the number of obstacles is 5, and the environment size is 20 by 20.}\label{fig:differentSolvers}
\end{figure}

We find our results at once encouraging and discouraging.  On the one hand, we have to wait more than 13 minutes for the solver to synthesize a 20 step motion plan, a substantial time.  On the other hand, as recently as 2014, before the enumerative solver fully supported let expressions, we would have had to wait more than 16 minutes.  See Figure \ref{fig:differentSolvers} for a comparison of the synthesis times for one of our benchmarks using the 2014 {\sc Stochastic} solver, the 2014 {\sc Enumerative} solver, and the 2015 {\sc Enumerative} solver.  Even on a length 20 motion plan, the older solver is more than 24\% slower.  For longer motion plans, the difference in performance is even more drastic.  In these early years of the SyGuS formalism, tools for solving SyGuS problems are rapidly improving.  As they continue to develop, we expect that synthesis times for this approach will continue to drop, making approaches like ours increasingly feasible.

Of our scalability results, the AST depth results are certainly the most discouraging.  Search space explosion is an inherent and inevitable feature of a SyGuS-based approach.  With each increase in depth, the solver must of course search a much larger space of programs.  Still, even the AST depth results leave us hopeful.  Although the complexity of the output program clearly has a tremendous impact on the feasibility of using our approach, our experience designing the benchmarks for the AST depth experiment left us surprised by the number of problems that can be solved with small programs.  For the AST depth experiment, we needed to identify a benchmark that could only be solved with a program of depth 5 or greater.  To do this, we designed tasks for which we ourselves would have written programs with depth 5 ASTs.  To our surprise, several of these tasks were solved by the synthesizer using programs with depth 4 ASTs before we ultimately created a variation for which the synthesizer was unable to find a smaller form.  Without any evidence of whether real world problems would behave similarly, we cannot be sure how encouraging we should find this experience.  We expect to explore this in future work.  We are hopeful, however, that applying synthesis to real world problems will also produce surprising and surprisingly small programs.

Given that our tool offers a synthesize-once approach, we are reasonably satisfied with the current synthesis times.  We imagine that -- in light of how easy it is to express a problem instance for our tool -- users might be willing to express a motion planning problem with our system and let it run a few minutes, or even overnight, before attempting to write an appropriate reactive motion plan by hand.  It is worth noting that the developers of existing SMT-based motion planning tools are pleased to see synthesis times even into the 4 hour range \cite{saha2014,saha2016}.   Ultimately, since this tool offers reactivity without having to replan during execution, the synthesis time is truly a one-time cost.  If users need a provably correct motion plan, they may be willing to accept some waiting at design time.

In short, although we would certainly like to see better performance results, we do believe SyGuS-based reactive motion planning is a feasible direction.  Taking a synthesize-once approach, rather than the alternative replanning approach, substantially reduces the time pressure.  Even more importantly, we expect to continue to see advances in SyGuS solver technology.  Because we use the standardized SyGuS format to express our problem, we can plug any current or future SyGuS solver into our tool.  Thus, our technique can be easily upgraded as the state of the art in SyGuS solving advances.

\section{Related Work}

In this section we discuss the classes of tools most closely related to our own work.  In particular, we break this work into two primary categories: tools that use existing functional synthesis formalisms for synthesizing non-reactive motion planners and tools that use custom or non-functional synthesizers to produce reactive motion planners.

\paragraph{Synthesis of Non-Reactive Motion Plans.}

Although we do not know of any tools that synthesize reactive motion plans by compiling directly to functional solvers, there are certainly tools that use off-the-shelf functional solvers to synthesize non-reactive motion plans.  We have already mentioned Complan \cite{saha2014}, an SMT-based motion planner for multi-robot systems.  We follow in Complan's footsteps in translating motion planning problems into existing synthesis formalisms, discretizing the space, and in decoupling the production of motion primitives from the decisions about when to use them.  A new tool from the same group, Implan \cite{saha2016}, takes the Complan approach even further through robot grouping, but still focuses on stable environments.

There has been a strong interest, especially recently, in tools that use SMT solvers to synthesize motion plans.  Like our tool, some focus on identifying a path between two points \cite{hungSMT}.  Others operate on novel input formats like Nedunuri's `program outlines', the input to SMT-based Robosynth \cite{nedunuri2014smt}.   The inputs and goals vary from tool to tool, but the heavy use of discretization and the clean compilation to existing functional solver technology are consistent.


\paragraph{Synthesis of Reactive Motion Plans.}

There is a growing body of work on synthesizing reactive motion planners, and on the custom algorithms and approaches developed to support them.  These tools sometimes focus on motion primitives in addition to the higher-level motion plans that use them \cite{LivingstonM, motionPrimitives}.  They often resynthesize throughout the process of execution \cite{recedingHorizon,yoshida}; sometimes they resynthesize to make use of increased environment knowledge during execution, and sometimes to achieve scalability by breaking the planning problem into smaller subproblems (following a receding horizon model).  Some combine design-time planning with online planning \cite{reactive}.  In short, there is a great variety of tools for generating reactive motion plans.  They also utilize a great variety of algorithms for handling their synthesis challenges.


There are also many tools that use reactive solvers to directly solve reactive motion planning problems.  Typically these tools frame their motion planning problems as games between the robot and an adversarial environment and use synthesizers that accept specifications from the generalized reactivity(1) fragment of LTL.  SmalL bUt Complete GROne Synthesizer (Slugs) \cite{slugs} and gr1c \cite{gr1c} are two such synthesizers.  Both produce automata from generalized reactivity(1) (GR(1)) specifications.  With the recent strong interest in high-level specification formats for reactive synthesis \cite{specificationFormat, TLSF, multiParadigm} and the introduction of a reactive synthesis competition SYNTCOMP \cite{syntcomp}, it seems likely stand-alone reactive synthesizers will become increasingly available and usable.  

Although we do not know of any reactive motion planners that use these higher-level interfaces to reactive synthesizers, there are certainly techniques that use GR(1) specifications and existing off-the-shelf reactive synthesis tools.  As an illustrative example, we look in more detail at the Kress-Gazit et al. tool presented in \cite{temporalLogic}.  Kress-Gazit et al. break their reactive motion plans into two parts, a discrete component that produces the high-level plan and a continuous component that implements the plan.  They produce the discrete component using the Piterman et al. algorithm presented in \cite{srd}.  The Piterman et al. algorithm takes GR(1) specifications as input and, if the input specification is realizeable, produces an automaton which offers one possible implementation of the system player.  The Kress-Gazit et al. tool accepts specifications of the same form, treats the robot and its environment as the two players, and uses the output automaton to directly determine the robot's motion plan.  Although these reactive synthesizers may offer motion planning tool builders less control over the form of the output motion plans, they certainly do offer off-the-shelf techniques that can be cleanly applied to reactive motion planning problems.  

Overall, the motion planners that use reactive solvers produce effective plans but offer the planners' builders less control over the plans.  New developments in reactive synthesis indicate that this may be changing as the reactive synthesis field moves towards giving users more control over the form of their output automata.  For instance, bounded synthesis \cite{boundedSynthesis} allows users to place limits on the number of states in the output automata.  While this does not reach the level of control that a developer can achieve with a functional solver, it offers an alternative to the unreadable automata with tens of thousands of states that are common outputs for standard reactive solvers.  If this progress continues, we expect that the advantages of a functional over a reactive approach to motion planning may diminish.  We believe the primary benefits of using a functional solver are control over the form of the output motion plans, the opportunity to easily explore many different output spaces (grammars), the opportunity to restrict the output space based on domain knowledge, and the potential to produce readable and adaptable motion plans.  We hope that as reactive solvers continue to develop, they will acquire many of the same advantages.

   
We know of one work that comes close to using a functional solver to synthesize reactive motion plans, Juniwal's {\sc Sketch}-based approach \cite{juniwal}.  Although this work does not synthesize complete motion plans, it does help users complete motion plans that they have already partially implemented.  The inputs to this approach are an LTL specification and a program with holes.  It uses the functional {\sc Sketch} synthesizer to fill the holes, then applies a verifier to the result.  If, for a given partial program, {\sc Sketch} cannot find a feasible candidate, the user may write a new partial program and try again.  If this partial program suggestion loop were automated (essentially building up a SyGuS-like layer on top of {\sc Sketch} \footnote{Note that indeed one SyGuS solver is implemented on top of {\sc Sketch}.}), this approach could be used to synthesize whole reactive motion plans.

\paragraph{High-Level Synthesis Tools Compiling to SyGuS.}

To our knowledge, no other high-level synthesis tool has yet attempted to compile to SyGuS.  

\section{Conclusion}

In environments with adversarial or moving obstacles, it is often valuable to allow robots to react to their surroundings.  While past synthesis tools have used off-the-shelf functional solvers to produce non-reactive motion plans, our tool is the first to use a functional solver to directly synthesize reactive motion plans.  By compiling motion planning problems to the new SyGuS formalism, we can synthesize not just a straight-line sequence of actions, but control flow for picking actions based on a robot's current status.  We have evaluated the performance of our approach over a range of program sizes, and the results indicate that a SyGuS-based approach is viable.  We believe the success of this approach offers an indication that other high-level synthesis goals -- and reachability games in particular -- which would previously have demanded the development of custom algorithms and solvers may now be in reach of sophisticated synthesis formalisms like SyGuS.

\section*{Acknowledgements}

We would like to thank Sanjit A. Seshia for suggesting this problem and for his thoughtful and insightful advice.  We also thank Indranil Saha, for detailed explanations of his work; Abhishek Udupa and Mukund Raghothaman, for the use of their SyGuS solvers; and the EECS 219C course at UC Berkeley.



\bibliographystyle{eptcs}
\bibliography{refs}

\begin{appendix}
\renewcommand{\thesection}{\Alph{section}}

\section{Full SyGus Specification}
\label{appendix:fullSpec}
\begin{lstlisting}[caption={The full SyGuS specification generated for the sample problem from Section 1.},label={lst:constraint},basicstyle=\tiny,breaklines=true]
(set-logic LIA)


(define-fun get-y ((currPoint Int)) Int 
(ite (< currPoint 5) 0 (ite (< currPoint 10) 1 (ite (< currPoint 15) 2 (ite (< currPoint 20) 3 (ite (< currPoint 25) 4 (ite (< currPoint 30) 5 6)))))))

(define-fun get-x ((currPoint Int)) Int
	(- currPoint (* (get-y currPoint) 5)))

(define-fun interpret-move (( currPoint Int ) ( move Int)) Int
(ite (= move 0) currPoint 
(ite (= move 1)  (ite (or (< (+ (get-y currPoint) 1) 0) (>= (+ (get-y currPoint) 1) 7))     currPoint (+ currPoint  5)) 
(ite (= move 2)  (ite (or (< (+ (get-x currPoint) 1) 0) (>= (+ (get-x currPoint) 1) 5))     currPoint (+ currPoint  1)) 
(ite (= move 3)  (ite (or (< (+ (get-y currPoint) -1) 0) (>= (+ (get-y currPoint) -1) 7))     currPoint (+ currPoint  -5)) 
(ite (= move 4)  (ite (or (< (+ (get-x currPoint) -1) 0) (>= (+ (get-x currPoint) -1) 5))     currPoint (+ currPoint  -1)) 
currPoint))))))

(define-fun interpret-move-obstacle-0 (( currPoint Int ) ( move Int)) Int
(ite (= move 0)  (ite (or (< (+ (get-y currPoint) 1) 0) (>= (+ (get-y currPoint) 1) 7))     currPoint (+ currPoint  5)) 
(ite (= move 1)  (ite (or (< (+ (get-y currPoint) -1) 0) (>= (+ (get-y currPoint) -1) 7))     currPoint (+ currPoint  -5)) 
currPoint)))

(define-fun interpret-move-obstacle-1 (( currPoint Int ) ( move Int)) Int
(ite (= move 0)  (ite (or (< (+ (get-y currPoint) 1) 0) (>= (+ (get-y currPoint) 1) 7))     currPoint (+ currPoint  5)) 
(ite (= move 1)  (ite (or (< (+ (get-y currPoint) -1) 0) (>= (+ (get-y currPoint) -1) 7))     currPoint (+ currPoint  -5)) 
currPoint)))

(define-fun no-overlap-one-move-combination-2-2 ((p0 Int) (p1 Int) (p2 Int) (p3 Int)) Bool
	(and (not (= p0 p2)) (and (not (= p0 p3)) (and (not (= p1 p2)) (not (= p1 p3))))))

(define-fun no-overlaps-0 (( currPoint Int ) ( move Int) (obstacleCurrPoint Int) (obstacleMove Int)) Bool
	(= 1
	(ite (= move 0) 
		(ite (= obstacleMove 0) (ite (no-overlap-one-move-combination-2-2 currPoint (+ (+ currPoint 0) 0) obstacleCurrPoint (+ (+ obstacleCurrPoint 0) 5)) 1 0)
		(ite (= obstacleMove 1) (ite (no-overlap-one-move-combination-2-2 currPoint (+ (+ currPoint 0) 0) obstacleCurrPoint (+ (+ obstacleCurrPoint 0) -5)) 1 0) 0))
	(ite (= move 1) 
		(ite (= obstacleMove 0) (ite (no-overlap-one-move-combination-2-2 currPoint (+ (+ currPoint 0) 5) obstacleCurrPoint (+ (+ obstacleCurrPoint 0) 5)) 1 0)
		(ite (= obstacleMove 1) (ite (no-overlap-one-move-combination-2-2 currPoint (+ (+ currPoint 0) 5) obstacleCurrPoint (+ (+ obstacleCurrPoint 0) -5)) 1 0) 0))
	(ite (= move 2) 
		(ite (= obstacleMove 0) (ite (no-overlap-one-move-combination-2-2 currPoint (+ (+ currPoint 1) 0) obstacleCurrPoint (+ (+ obstacleCurrPoint 0) 5)) 1 0)
		(ite (= obstacleMove 1) (ite (no-overlap-one-move-combination-2-2 currPoint (+ (+ currPoint 1) 0) obstacleCurrPoint (+ (+ obstacleCurrPoint 0) -5)) 1 0) 0))
	(ite (= move 3) 
		(ite (= obstacleMove 0) (ite (no-overlap-one-move-combination-2-2 currPoint (+ (+ currPoint 0) -5) obstacleCurrPoint (+ (+ obstacleCurrPoint 0) 5)) 1 0)
		(ite (= obstacleMove 1) (ite (no-overlap-one-move-combination-2-2 currPoint (+ (+ currPoint 0) -5) obstacleCurrPoint (+ (+ obstacleCurrPoint 0) -5)) 1 0) 0))
	(ite (= move 4) 
		(ite (= obstacleMove 0) (ite (no-overlap-one-move-combination-2-2 currPoint (+ (+ currPoint -1) 0) obstacleCurrPoint (+ (+ obstacleCurrPoint 0) 5)) 1 0)
		(ite (= obstacleMove 1) (ite (no-overlap-one-move-combination-2-2 currPoint (+ (+ currPoint -1) 0) obstacleCurrPoint (+ (+ obstacleCurrPoint 0) -5)) 1 0) 0)) 0)))))))

(define-fun no-overlaps-1 (( currPoint Int ) ( move Int) (obstacleCurrPoint Int) (obstacleMove Int)) Bool
	(= 1
	(ite (= move 0) 
		(ite (= obstacleMove 0) (ite (no-overlap-one-move-combination-2-2 currPoint (+ (+ currPoint 0) 0) obstacleCurrPoint (+ (+ obstacleCurrPoint 0) 5)) 1 0)
		(ite (= obstacleMove 1) (ite (no-overlap-one-move-combination-2-2 currPoint (+ (+ currPoint 0) 0) obstacleCurrPoint (+ (+ obstacleCurrPoint 0) -5)) 1 0) 0))
	(ite (= move 1) 
		(ite (= obstacleMove 0) (ite (no-overlap-one-move-combination-2-2 currPoint (+ (+ currPoint 0) 5) obstacleCurrPoint (+ (+ obstacleCurrPoint 0) 5)) 1 0)
		(ite (= obstacleMove 1) (ite (no-overlap-one-move-combination-2-2 currPoint (+ (+ currPoint 0) 5) obstacleCurrPoint (+ (+ obstacleCurrPoint 0) -5)) 1 0) 0))
	(ite (= move 2) 
		(ite (= obstacleMove 0) (ite (no-overlap-one-move-combination-2-2 currPoint (+ (+ currPoint 1) 0) obstacleCurrPoint (+ (+ obstacleCurrPoint 0) 5)) 1 0)
		(ite (= obstacleMove 1) (ite (no-overlap-one-move-combination-2-2 currPoint (+ (+ currPoint 1) 0) obstacleCurrPoint (+ (+ obstacleCurrPoint 0) -5)) 1 0) 0))
	(ite (= move 3) 
		(ite (= obstacleMove 0) (ite (no-overlap-one-move-combination-2-2 currPoint (+ (+ currPoint 0) -5) obstacleCurrPoint (+ (+ obstacleCurrPoint 0) 5)) 1 0)
		(ite (= obstacleMove 1) (ite (no-overlap-one-move-combination-2-2 currPoint (+ (+ currPoint 0) -5) obstacleCurrPoint (+ (+ obstacleCurrPoint 0) -5)) 1 0) 0))
	(ite (= move 4) 
		(ite (= obstacleMove 0) (ite (no-overlap-one-move-combination-2-2 currPoint (+ (+ currPoint -1) 0) obstacleCurrPoint (+ (+ obstacleCurrPoint 0) 5)) 1 0)
		(ite (= obstacleMove 1) (ite (no-overlap-one-move-combination-2-2 currPoint (+ (+ currPoint -1) 0) obstacleCurrPoint (+ (+ obstacleCurrPoint 0) -5)) 1 0) 0)) 0)))))))

(define-fun no-overlaps-one-step ((currPoint Int) (move Int) (o0pos Int) (o0move Int) (o1pos Int) (o1move Int)) Bool
	(and (no-overlaps-0 currPoint move o0pos o0move) (no-overlaps-1 currPoint move o1pos o1move)))



(declare-var o0-mov0 Int)
(declare-var o0-mov1 Int)
(declare-var o0-mov2 Int)
(declare-var o0-mov3 Int)
(declare-var o0-mov4 Int)
(declare-var o0-mov5 Int)
(declare-var o1-mov0 Int)
(declare-var o1-mov1 Int)
(declare-var o1-mov2 Int)
(declare-var o1-mov3 Int)
(declare-var o1-mov4 Int)
(declare-var o1-mov5 Int)

(synth-fun move ((currPoint Int) (o0 Int) (o1 Int)) Int
	((Start Int (
		MoveId
		(ite StartBool Start Start)))
    (MoveId Int (0
		1
		2
		3
		4
  	))
	(CondInt Int (
		(get-y currPoint) ;y coord
		(get-x currPoint) ;x coord
		(get-y o0)
		(get-x o0)
		(get-y o1)
		(get-x o1)
		(+ CondInt CondInt)
		(- CondInt CondInt)
		-1
		0
		1
		2
		3
		4
		5
		6
				))
	(StartBool Bool ((and StartBool StartBool)
		(or  StartBool StartBool)
		(not StartBool)
		(<=  CondInt CondInt)
		(=   CondInt CondInt))))) 
 
 (constraint
	(or
		(not (and (or (= o0-mov0 0) (= o0-mov0 1)) (and (or (= o0-mov1 0) (= o0-mov1 1)) (and (or (= o0-mov2 0) (= o0-mov2 1)) (and (or (= o0-mov3 0) (= o0-mov3 1)) (and (or (= o0-mov4 0) (= o0-mov4 1)) (and (or (= o0-mov5 0) (= o0-mov5 1)) (and (or (= o1-mov0 0) (= o1-mov0 1)) (and (or (= o1-mov1 0) (= o1-mov1 1)) (and (or (= o1-mov2 0) (= o1-mov2 1)) (and (or (= o1-mov3 0) (= o1-mov3 1)) (and (or (= o1-mov4 0) (= o1-mov4 1)) (or (= o1-mov5 0) (= o1-mov5 1))))))))))))))

	 (let ( (o0-pos0 Int 27) (o1-pos0 Int 28)) (let ( (o0-pos1 Int (interpret-move-obstacle-0 o0-pos0 o0-mov0)) (o1-pos1 Int (interpret-move-obstacle-1 o1-pos0 o1-mov0))) (let ( (o0-pos2 Int (interpret-move-obstacle-0 o0-pos1 o0-mov1)) (o1-pos2 Int (interpret-move-obstacle-1 o1-pos1 o1-mov1))) (let ( (o0-pos3 Int (interpret-move-obstacle-0 o0-pos2 o0-mov2)) (o1-pos3 Int (interpret-move-obstacle-1 o1-pos2 o1-mov2))) (let ( (o0-pos4 Int (interpret-move-obstacle-0 o0-pos3 o0-mov3)) (o1-pos4 Int (interpret-move-obstacle-1 o1-pos3 o1-mov3))) (let ( (o0-pos5 Int (interpret-move-obstacle-0 o0-pos4 o0-mov4)) (o1-pos5 Int (interpret-move-obstacle-1 o1-pos4 o1-mov4)))
 (let ((pos0 Int 6)) (let ((mov0 Int (move pos0 27 28))) (let ((pos1 Int (interpret-move pos0 mov0))) (let ((mov1 Int (move pos1 o0-pos1 o1-pos1))) (let ((pos2 Int (interpret-move pos1 mov1))) (let ((mov2 Int (move pos2 o0-pos2 o1-pos2))) (let ((pos3 Int (interpret-move pos2 mov2))) (let ((mov3 Int (move pos3 o0-pos3 o1-pos3))) (let ((pos4 Int (interpret-move pos3 mov3))) (let ((mov4 Int (move pos4 o0-pos4 o1-pos4))) (let ((pos5 Int (interpret-move pos4 mov4))) (let ((mov5 Int (move pos5 o0-pos5 o1-pos5))) (let ((pos6 Int (interpret-move pos5 mov5)))

	(and
		(= pos6 24)
		(and (no-overlaps-one-step pos0 mov0 27 o0-mov0 28 o1-mov0) (and (no-overlaps-one-step pos1 mov1 o0-pos1 o0-mov1 o1-pos1 o1-mov1) (and (no-overlaps-one-step pos2 mov2 o0-pos2 o0-mov2 o1-pos2 o1-mov2) (and (no-overlaps-one-step pos3 mov3 o0-pos3 o0-mov3 o1-pos3 o1-mov3) (and (no-overlaps-one-step pos4 mov4 o0-pos4 o0-mov4 o1-pos4 o1-mov4) (no-overlaps-one-step pos5 mov5 o0-pos5 o0-mov5 o1-pos5 o1-mov5))))))))))))))))))))))))))))

(check-synth)
\end{lstlisting}
\end{appendix}

\end{document}